\documentclass{Rinton-P9x6}
\begin{document}
\title{Fundamental limitation on qubit operations due to the Bloch-Siegert Oscillation }
\author{M. S. Shahriar   and  Prabhakar Pradhan }
\address{ Department  of Electrical and Computer Engineering,  Northwestern Univeristy \\
                 Evanston, IL 60208  and \\
     Research Laboratory of Electronics,  Massachusetts Institute of Technology \\ 
                  Cambridge, MA 02139 }
\maketitle
\abstracts{
We show that 
if the Rabi frequency is comparable to the Bohr frequency so that the rotating wave 
approximation is inappropriate, an extra oscillation is present with the Rabi oscillation. 
We discuss how the sensitivity of the degree of excitation to the phase of the field 
may pose severe constraints on precise rotations of quantum bits involving low-frequency 
transitions. We present a scheme for observing this effect in an atomic beam.}

It is well known that the amplitude of an atomic state is necessarily complex. 
The electric or magnetic field generated by an oscillator, on the other hand, is real, composed of the 
sum of two complex components.  In describing semiclassically the atom-field interaction involving
such a field, one often side-steps this difference by making the so-called  rotating wave 
approximation (RWA), under which only one of the two complex components is kept, and the counter-rotating 
part is ignored.  
%
We discuss how the sensitivity of the degree of excitation to the phase of the field 
poses severe constraints on precise rotations of quantum 
bits involving low-frequency transitions. We also present a scheme for observing this effect in an atomic beam.

 We consider an ideal two-level system  where  a ground state $|0\rangle$ 
is coupled to an excited state $|1\rangle $, and  the 
$0 \leftrightarrow 1$ transitions are magnetic dipolar, 
with a transition frequency $\omega$, and the  magnetic field
is  of the form $B=B_0 \cos(\omega t+\phi)$. 
We now summarize briefly the two-level dynamics without the RWA.
In the dipole approximation, the Hamiltonian can be written as:
\begin{equation} 
\hat{H} = \epsilon ( \sigma_0 -\sigma_z )/2 + g(t) \sigma_x
\label{hmt_original}
\end{equation}
where $g(t) = -g_0\left[\exp (i\omega t + i\phi)+c.c. \right] /2$, 
$\sigma_i$ are Pauli matrices, and  $\epsilon=\omega$ 
corresponds to resonant excitation. The state vector is written as:
\begin{equation} 
|\xi(t)\rangle = \left( \begin{array}
{c} C_0(t)  \\  C_1(t)
\end{array} \right). 
\label{ket_c_original}
\end{equation}
We perform  a rotating wave transformation by operating on  $|\xi(t)\rangle$
with the unitary operator $\hat{Q}$, where:                                                      
$\hat{Q} = (\sigma_0 + \sigma_z )/2 +  \exp (i\omega t + i\phi)(\sigma_0 - \sigma_z)/2 $. 
The Schr\"{o}dinger equation then  takes the form (setting $\hbar=1$):
$\dot{ |\tilde{\xi}\rangle } = -i \tilde{H(t)} |\tilde{\xi}(t) \rangle $ 
where the effective Hamiltonian is given by:
\begin{equation} 
\tilde{H} =   \alpha(t)\sigma_{+} + \alpha^{*}(t) \sigma_{-},
\label{hmt_tilde}
\end{equation}
with $\alpha(t) = - (g_0/2)\left[\exp (-i2\omega t- i2\phi)+1 \right] $, 
and in the rotating frame the state vector is:
\begin{equation} 
|\tilde{\xi}(t) \rangle  \equiv \hat{Q}| {\xi}(t)\rangle= \left( 
\begin{array}{c} \tilde{C}_0(t)  \\  \tilde{C}_1(t)
\end{array} \right). 
\label{ket_c_tilde}
\end{equation}
Now, by  making  the rotating wave approximation (RWA),
corresponding to dropping the fast oscillating term in $\alpha(t)$ 
one  ignores  effects (such as the Bloch-Siegert shift ) of the 
order of  $(g_0/\omega)$, which can easily be observed in experiment if 
$g_0$ is large \cite{Corney77,Bloch40}.
Otherwise,  by  choosing $g_0$ to  be small enough, one can make 
the RWA for any value of $\omega$.  We explore here  both regimes, 
and we find the general results without the RWA.

From Eqs.\ref{hmt_tilde} and \ref{ket_c_tilde}, we get two  coupled differential  equations:
%
%
\begin{eqnarray}
\dot{\tilde{C}}_0(t) & = & i(g_0/2)\left[ 1+ \exp  (-i2\omega t-i2\phi)\right 
]\tilde{C}_1(t)\\
\label{c_diff_eqn_a}
\dot{\tilde{C}}_1(t) & = & i(g_0/2)\left[ 1+ \exp  (+i2\omega t+i2\phi)\right] 
\tilde{C}_0(t).
\label{c_diff_eqn_b}
\end{eqnarray}
%
%
Let  $|C_0(0)|^2=1$ be the initial condition, and we proceed 
to find  approximate analytical solutions of 
Eqs.5 and 6. 
Due to the periodic nature of the effective Hamiltonian, the general 
solution to Eqs.5 and 6 
can be written in the form:                                                
\begin{equation} 
|\tilde{\xi}(t)\rangle = \sum_{n=-\infty}^{\infty}
\left( \begin{array}{c} a_n  \\  b_n \end{array} \right) \exp(  n(-i2\omega t-i2\phi)). 
\label{xi_Bloch_expn}
\end{equation}
Inserting Eq.\ref{xi_Bloch_expn}  in  Eqs.5 and 6 
we get for all $n$ :
\begin{eqnarray}
  \dot{a}_n & = & i2n\omega a_n + i g_0(b_n +b_{n-1})/2, \\
  \dot{b}_n & = & i2n\omega b_n + i g_0(a_n +a_{n+1})/2.
\end{eqnarray}
\label{a_n_b_n_eqn}
%
%
Here, the coupling between $a_0$ and $b_0$ is the conventional 
one  when the RWA is made. The couplings to the nearest 
neighbors, $a_{\pm 1}$ and $b_{\pm 1}$, are detuned by an amount 
$2\omega$, and so on.  To the lowest order in $(g_0/\omega)$, we can ignore 
terms with $|n|>1$, thus yielding a truncated set of six equations for 
$\dot{a}_0,\dot{b}_0,\dot{a}_{\pm 1},\dot{b}_{\pm 1}$.                
Now consider $g_0$ to have a time-dependence of the form 
$g_0(t)=g_{0M}\left[1-\exp(-t/\tau_{sw})\right]$, 
where the switching time constant  $\tau_{sw}$
is large compared to other characteristic time scales such as $1/\omega$  and $1/g_{0M}$.  
Then, these equations can be solved  by the method of adiabatic 
elimination, which is  valid to first order in $\eta\equiv(g_0/4\omega )$. 
(Note  $\eta(t)=\eta_0\left[1-\exp(-t/\tau_{sw})\right]$, 
where $\eta_0\equiv(g_0M/4\omega) )$. 
To solve the set of equations above, we consider first $a_{-1}$ and $b_{-1}$. 
Define $\mu_{\pm}\equiv (a_{-1}\pm b_{-1})$, then one can write
$\dot{\mu}_{\pm}  =  -i(2\omega+ g_0/2 ) \mu_{\pm} \pm i g_0 a_{0}/2 $. 
Adiabatic following then yields (again, to lowest order in $\eta$ ):
$a_{-1} \approx 0 $ and  $b_{-1} \approx \eta a_{0}$.   
Likewise, we can show that $a_{1} \approx -\eta b_0$ and $b_{1} \approx 0$.   
%
%
The only non-vanishing  (to lowest order in $\eta $ with $|C_0(t=0)|^2=1$) 
terms in the solution of Eqs.5 and 6  are:
%
%
\begin{eqnarray}
C_0(t) &=& \cos(g_0'(t)t/2) - i\eta e^{-i(2\omega t +2\phi)} \sin(g_0'(t)t/2), \\
C_1(t) &=& i e^{-i(\omega t + \phi)} [ \sin(g_0'(t)t/2)   
        - i\eta e^{+i(2\omega t +2\phi)} \cos(g_0'(t)t/2) ],
\end{eqnarray}
\label{c_total_eqn}
where $g_0'(t)=1/t \int_0^t g_{0}(t)dt = g_0 \left[1-(t/\tau_{sw})^{-1}\exp(-t/\tau_{sw})\right].$
 
 To lowest order in $\eta$, this solution is normalized at all times. 
Note that if one wants to carry this excitation on an ensemble 
of atoms using a $\pi/2$  pulse and measure the population of the 
state  $|1\rangle$  after the excitation terminates (at $t=\tau $ when 
$g'(\tau)\tau/2= \pi /2$ ), the output signal will be:
\begin{equation}
|C_{1}(g_0'(\tau),\phi)|^2 =\frac{1}{2}\left[1+2\eta 
\sin(2\omega\tau + 2\phi)\right],
\label{population_c1}
\end{equation}
which contains information of both the  amplitude and the phase of the 
driving  field $B$. This clearly indicates that the Rabi transition probability 
depends on the total phase  $\phi_{\tau}= \omega\tau + \phi$ of the driving field.\\
\begin{figure}[!htb]
\centering
\includegraphics[scale=0.3]{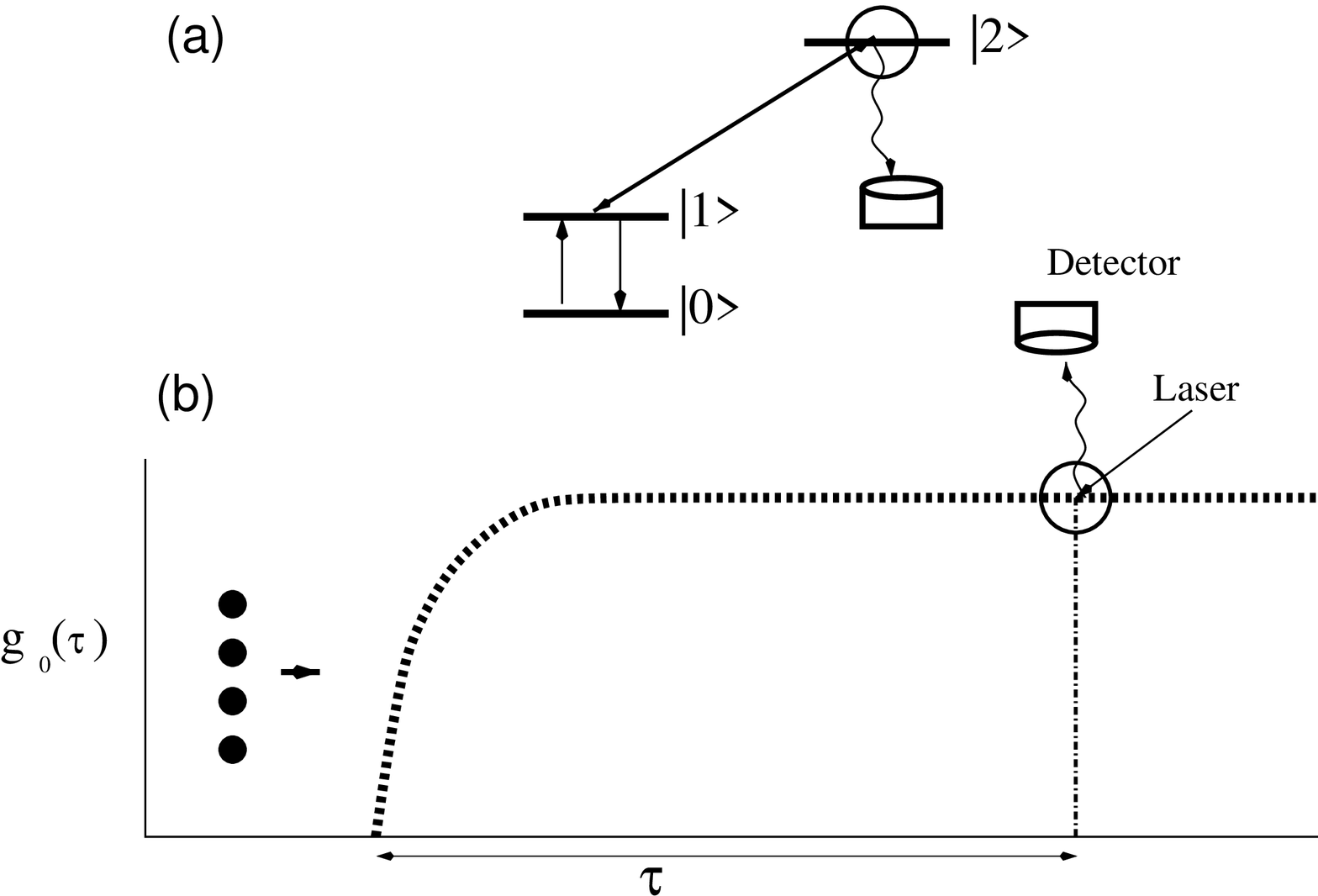}
\includegraphics[scale=0.4]{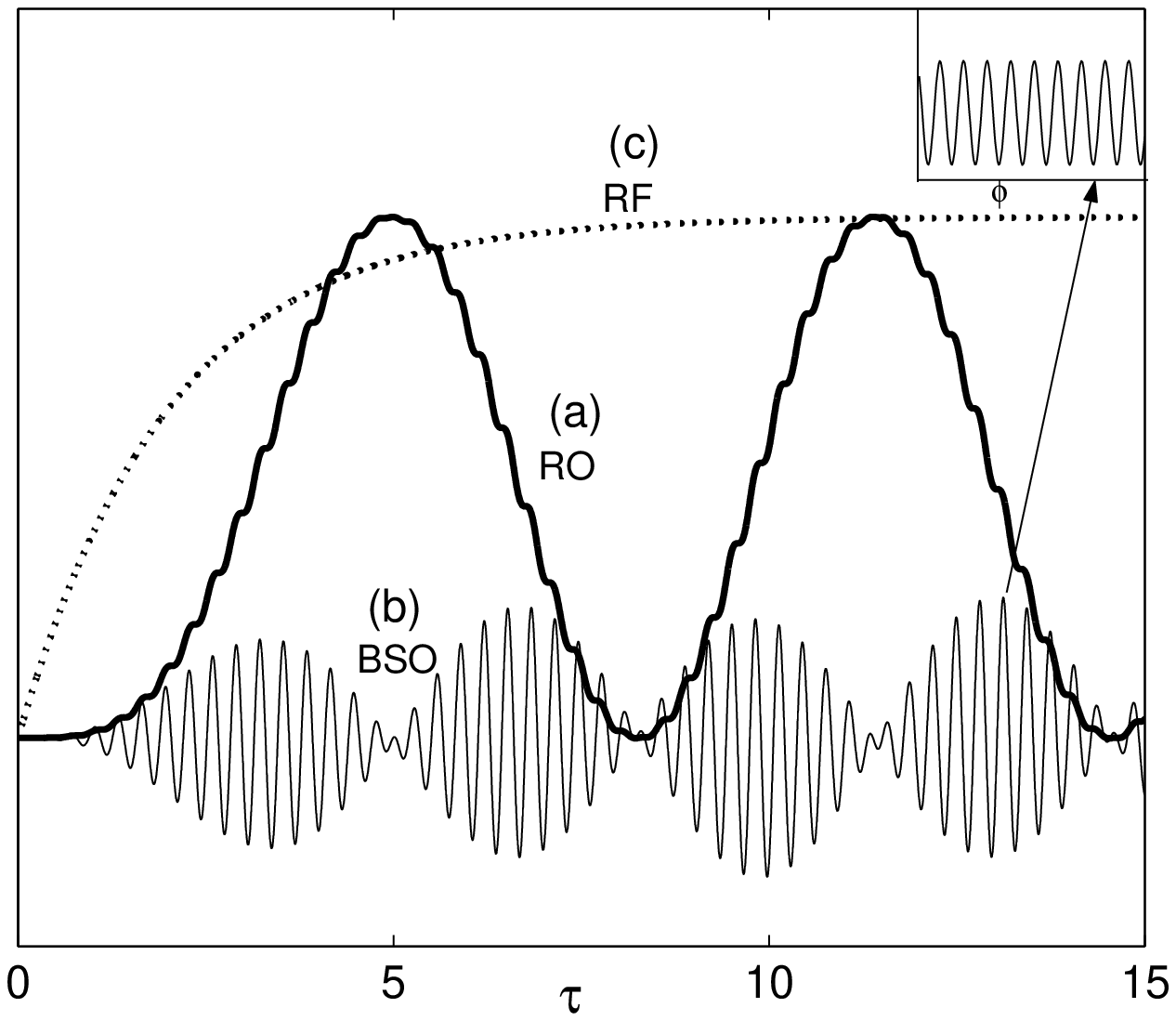}
\caption{ 
Left: Schematic illustration of an experimental arrangement for measuring the phase 
dependence of the population of the excited state $|1\rangle$: 
(a) The microwave field couples the ground state 
($|0\rangle$) to the excited state ($|1\rangle$).  
A third level, $|2\rangle$, which can be coupled 
to $|1\rangle$ optically, is  used to measure the 
population of $|1\rangle$ via fluorescence detection. 
(b) The microwave field is  turned on adiabatically with a switching 
time-constant $\tau_{sw}$, and the fluorescence is monitored after a total 
interaction time of $\tau$. 
 Right: Illustration of the Bloch-Siegert Oscillation (BSO):  (a)  The population of state 
$|1\rangle$, as a function of the interaction time $\tau$ , showing the BSO superimposed on the 
conventional Rabi oscillation. (b) The BSO oscillation (amplified scale) by itself, 
produced by subtracting  the Rabi oscillation from the plot in (a). 
(c) The time-dependence of the Rabi frequency.
Inset: BSO as a function of the absolute phase of the field with fixed $\tau$.
}
\end{figure}
A physical realization of this result can be best appreciated  by
considering an experimental arrangement of the type illustrated in Fig.1(left).
And plot of the Rabi oscillations is shown in Fig.1(right).
Under the RWA, the curve of Fig.1(a)(right) represents the conventional Rabi 
oscillation. However, we notice here some additional oscillation, which is magnified 
and shown separately in Fig.1(b)(right), produced by subtracting the 
conventional Rabi oscillation ($\sin^2(g(t)/2)$) from Fig.1(a)(right).
That is, Fig.1(b)(right) corresponds to what we call the  Bloch-Siegert Oscillation 
(BSO), given by  $\eta \sin(g'_0(\tau)\tau)\sin(2\phi_{\tau})$.
These analytical results agree very closely  to the results which 
are obtained via direct numerical integration of  Eqs.5 and 6. 
Note that the BSO is at twice the frequency of the driving field, and its amplitude is 
enveloped by a function that vanishes when all the atoms are in a single state.
This osillation will be  stonger  when the ratio of the resonance frequency to the Rabi 
frequency is large \cite{Bouwmeester00,Steane97a,Jonathan92}.
One should keep track of the phase of the excitation  field at 
the location of the qubit \cite{Pradhan02a,Pradhan02b} for low energy 
qubit systems \cite{Thomas82,Shahriar90,Shahriar97} with fast driving.

In conclusion, we have shown that when a two-level atomic system is driven by a strong periodic 
field, the Rabi oscillation is accompanied by another oscillation at twice the transition frequency.
This extra oscillation can limit the qubit operations in Rabi flopping without a proper matching of
the parameters. However, it has been shown that, this phase has potential application in 
teleportation.  By making use of distant entanglement, this mechanism 
may enable teleportation of the phase of a field that is encoded in the atomic 
state amplitude, and has potential applications to remote frequency locking 
\cite{Jozsa00,Lloyd01,Levy87,Shahriar01}.


We wish to acknowledge support from DARPA grant No. F30602-01-2-0546 under the QUIST
program, ARO grant No. DAAD19-001-0177 under the MURI program, 
and NRO grant No. NRO-000-00-C-0158.

\end{document}